\documentclass[12pt]{iopart}

\usepackage{graphicx}
\usepackage{bm}% bold math
\usepackage{color}
\begin{document}

\title{Bethe-lattice calculations for the phase diagram of a two-state Janus gas}

\author{Danilo B. Liarte}
\ead{danilo@if.usp.br}
\address{Institute of Physics, Caixa Postal 66318, CEP 05314-970, S\~ao Paulo, SP, Brazil}
\author{Silvio R. Salinas}
\ead{ssalinas@if.usp.br}
\address{Institute of Physics, Caixa Postal 66318, CEP 05314-970, S\~ao Paulo, SP, Brazil}

\vspace{10pt}
\begin{indented}
\item[]January 2015
\end{indented}

\begin{abstract}
We use a simple lattice statistical model to analyze the effects of
directional interactions on the phase diagram of a fluid of two-state Janus
particles. The problem is formulated in terms of nonlinear recursion relations
along the branches of a Cayley tree. Directional interactions are taken into
account by the geometry of this graph. Physical solutions on the Bethe lattice
(the deep interior of a Cayley tree) come from the analysis of the attractors
of the recursion relations. We investigate a number of situations, depending
on the concentrations of the types of Janus particles and the parameters of
the potential, and make contact with results from recent numerical simulations.
\end{abstract}

%\maketitle

\section{Introduction}

The production and characterization of solutions of Janus particles, whose
spherical surface is divided into hydrophobic and hydrophilic hemispheres,
have attracted the attention of a number of authors \cite{jiang08, gangwal10, reinhardt11, vissers13, jiang14, shin14}. Studies of the behavior of these colloidal systems must take into account that Janus particles interact in a different way depending on their relative orientations. A simple form of a pairwise directional potential has been proposed by Kern and Frenkel \cite{kern03}, who introduced a model of hard spheres with the addition of a square-well potential with directional attractive short-range interactions. The Kern-Frenkel model has been used in several analytical and numerical investigations (see e.g. \cite{sciortino09, sciortino10, vissers13, shin14}). In particular, a simplified two-state version of the Kern-Frenkel model, which is reminiscent of the Zwanzig approximation for liquid-crystalline models, and can be experimentally realised by the application of an electric field, has been extensively studied in \cite{maestre13, fantoni13}.

\begin{figure}[!ht]
\centering
\includegraphics[width=0.6\linewidth]{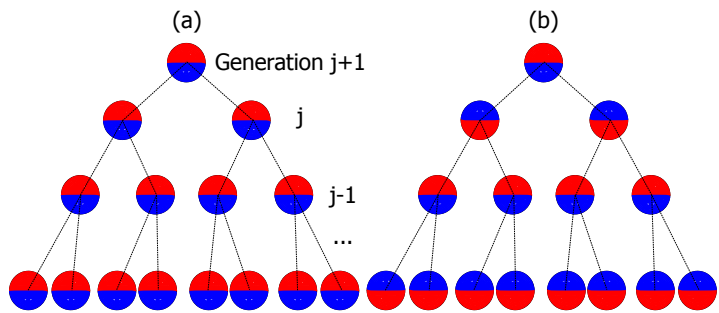}%
\caption{Sketch of some generations of a Cayley tree of ramification $r=2$.
The two-state system is formed by Janus particles of type $a$ (depicted as circles with red heads up in
this figure) and of type $b$ (blue heads up). The trees on the left (a), and on the right (b), illustrate macroscopic configurations with a high density of $a$-particles, and a modulated (``striped'') phase, respectively. \label{cayley}}
\end{figure}

These recent calculations provided the motivation to introduce a simple
lattice statistical model to analyze the effects of directional interactions
in a Janus gas of particles. It is well-known that lattice gas models on the
Bethe lattice, which is the deep interior of a large Cayley tree, lead to the
same (analytic) equations of state of the quasi-chemical approximation for a
simple lattice gas \cite{runnels67}. In the limit of infinite coordination
of the tree, it has been shown that one regains the usual mean-field solutions
\cite{thompson82, baxter82}. We then revisit the lattice gas problem
on a Cayley tree with the inclusion of directional interactions. We consider a
mixture of Janus particles of types $a$ and $b$, with hydrophobic
(hydrophilic) hemispheres in the upper (lower) half part of their respective
spherical surfaces, which amounts to considering a two-state, Ising-like,
representation of a Janus gas (see figure \ref{cayley}). In a grand canonical formulation, we add a
chemical potential to control the density of each type of particle. Taking
advantage of the geometrical structure of this tree, it becomes particularly
simple to introduce directional interactions between first-neighbor sites
along the branches of the graph. We choose the parameters of the potential,
and the concentration of the two types of particles, to make contact with the
available numerical simulations.

This article is organized as follows. In Section I, we describe the lattice
statistical model on a Cayley tree. We show that the solutions on the Bethe
lattice (deep in the interior of the tree) can be obtained from the analysis
of the recursion relations associated with a nonlinear discrete map. In
sections II and III, we analyze some particular cases, and make contact with
results from simulations. In particular, we show that attractive interactions
of the Kern-Frenkel form lead to layered modulated structures, which have
indeed been found in the simulations \cite{fantoni13}. Although we do not
make quantitative contacts with the simulations, we do claim to have used a
much simpler approach to obtain a number of analytical qualitatively results,
for a full range of model parameters. In particular, we provide a unified view
of the phase transitions described by a ``cubic diagram'' of interaction
parameters drawn by Fantoni and collaborators \cite{fantoni13}.

\section{Formulation of the problem on the Bethe Lattice}

We consider Janus particles restricted to two orientations. These particles
are represented by a set variables $\left\{  t_{i}\right\}  $, on the sites of
a tree, so that $t_{i}=1$ is associated with a particle of type $a$ on site
$i$, and $t_{i}=0$ represents a Janus particle of type $b$ on site $i$. The
pair interaction energy between nearest-neighbor sites $i$ and $j$ along the
branches of the tree is given by%
\begin{eqnarray}
E_{ij}=-(\epsilon_{aa}+\epsilon_{bb}-\epsilon_{ab}-\epsilon_{ba})t_{i}%
t_{j}-(\epsilon_{ab}-\epsilon_{bb})t_{i}
%\nonumber \\ && \quad
-(\epsilon_{ba}-\epsilon_{bb}%
)t_{j}-\epsilon_{bb}.
\end{eqnarray}
Note that $aa$ and $bb$ particles interact with energies $-\epsilon_{aa}$ and
$-\epsilon_{bb}$, respectively. Also, note that $ab$ and $ba$ particles
interact with (in general) different energies, $-\epsilon_{ab}\neq
-\epsilon_{ba}$. Of course, if $\epsilon_{ab}=\epsilon_{ba}$ we regain the
results for a usual lattice gas representation of a binary liquid mixture
\cite{runnels67}.%

In the appendix, we use standard treatments for a Cayley tree \cite{baxter82}, in order to obtain the recursion relations
\begin{equation}
\Xi_{j+1}^{a}=z\left[  e^{K_{aa}}\Xi_{j}^{a}+e^{K_{ab}}\Xi_{j}^{b}\right]
^{r}
\label{xi-a_recur}%
\end{equation}
and%
\begin{equation}
\Xi_{j+1}^{b}=\left[  e^{K_{ba}}\Xi_{j}^{a}+e^{K_{bb}}\Xi_{j}^{b}\right]
^{r},
\label{xi-b_recur}%
\end{equation}
for the partial grand partition function $\Xi_{j}^{a}$ ($\Xi_{j}^{b}$) that is associated with the sub-tree generated by a site at generation $j$ which is occupied by a particle of type $a$ ($b$). $K_{kl}=\beta\epsilon_{kl}$, for $k,l=a,b$, $\beta=1/k_{B}T$ is the inverse temperature, $z=\exp(\beta\mu)$ is the fugacity, and $\mu$ is the chemical potential (associated with particles of type $a$). It is now convenient to define the density of particles of type $a$ in
generation $j$,
\begin{equation}
\rho_{j}=\frac{\Xi_{j}^{a}}{\Xi_{j}^{a}+\Xi_{j}^{b}},
\end{equation}
so that Eqs. (\ref{xi-a_recur}) and (\ref{xi-b_recur}) may be written as a
single recursion relation%
\begin{equation}
\rho_{j+1}=f(\rho_{j}),\label{map-eq}%
\end{equation}
with
\begin{equation}
f(x)=\left\{  1+z\left[  \frac{e^{K_{ba}}x+e^{K_{bb}}(1-x)}{e^{K_{aa}%
}x+e^{K_{ab}}(1-x)}\right]  ^{r}\right\}  ^{-1},\label{map-eq2}%
\end{equation}
where $0\leq x\leq1$. This is the central result of this formulation. At this
point the problem is reduced to analyzing the general map given by Eq.
(\ref{map-eq}), from which we obtain the main features of the phase diagrams in
terms of temperature, $T=1/ (k_{B}\beta)$, and chemical potential $\mu$ (and with
different choices of the energy parameters).

It is interesting to investigate the limit of infinite coordination of the
tree, $r\rightarrow\infty$, with fixed values $r\epsilon_{aa}$, $r\epsilon
_{bb}$, $r\epsilon_{ab}$, and $r\epsilon_{ba}$. In this limit, it is easy to
show that
\begin{equation}
f(x)\rightarrow f_{\infty}(x)=\left(  1+e^{K_{1}+K_{2}\,x}\right)  ^{-1},
\end{equation}
where%
\begin{equation}
K_{1}=\beta\left(  \mu+\delta\right)  ,\quad K_{2}=\beta\Delta,
\end{equation}
with
\begin{equation}
\Delta=r\left(  \epsilon_{ab}+\epsilon_{ba}-\epsilon_{aa}-\epsilon
_{bb}\right)  ,\quad\delta=r\left(  \epsilon_{bb}-\epsilon_{ab}\right)  .
\end{equation}
This limit is known to lead to the solutions for an analogous fully-connected,
mean-field model \cite{thompson82, baxter82}. It is
important to remark that the phase diagrams depend on just two parameters,
$K_{1}$ and $K_{2}$, and that the parameter $\Delta$ plays a quite special role.

At high temperatures, both $f(x)$ and $f_{\infty}(x)$ tend to $1/2$, so that
$\rho\rightarrow1/2$ as $T\rightarrow\infty$. At low temperatures, there are
two possible scenarios depending on the sign of $\Delta$. If $\Delta<0$, there
is a discontinuous (first-order) line of transitions in a diagram in terms of
chemical potential and temperature. This border, which separates states with
low- and high-density of particles of type $a$, ends at a critical point. In
the second scenario, for $\Delta>0$, the analogous phase diagram displays a
critical line enclosing a cycle-$2$ periodic phase. This approach provides a
unified view of the phase transitions described by the choice of interaction
parameters according to the cubic diagram of Fantoni and collaborators (see table \ref{fantonis_table}). Except for the HS and SW models, which are non-interacting models in our lattice approach, all of the cases in this diagram are shown to fit into one of these two scenarios.
Also, although the finite-coordination map is more respectable than the
mean-field limit, we find no qualitative changes for trees with ramification
$r>1$ ($r=1$ corresponds to a one-dimensional model, which has no phase
transition at finite temperature). We will discuss these points in the
following sections.

\begin{table}[h]
\centering
\begin{tabular}{ | l | c c c c c | }
\hline
Model & $\epsilon_{aa}$ & $\epsilon_{ab}$ & $\epsilon_{ba}$ & $\epsilon_{bb}$  & $\Delta$ \\ \hline
HS & 0 & 0 & 0 & 0 & 0 \\
A0 & 0 & $\epsilon$ & 0 & 0 & $r \, \epsilon$ \\
I0 & $\epsilon$ & 0 & 0 & $\epsilon$ & $ - 2 \, r \, \epsilon$ \\
J0 & 0 & $\epsilon$ & $\epsilon$ & 0 & $2 \, r \, \epsilon$ \\
B0 & $\epsilon$ & $\epsilon$ & 0 & $\epsilon$ & $- r \, \epsilon$ \\
SW & $\epsilon$ & $\epsilon$ & $\epsilon$ & $\epsilon$ & 0 \\
\hline
\end{tabular}
\footnotesize
\caption{ Definition of the models and corresponding values of $\Delta$ according to nomenclature defined in \cite{fantoni13}. }
\label{fantonis_table}
\end{table}

\section{Discontinuous transition ($\Delta < 0$)}

\begin{figure}[!ht]
\centering
\includegraphics[width=0.6\linewidth]{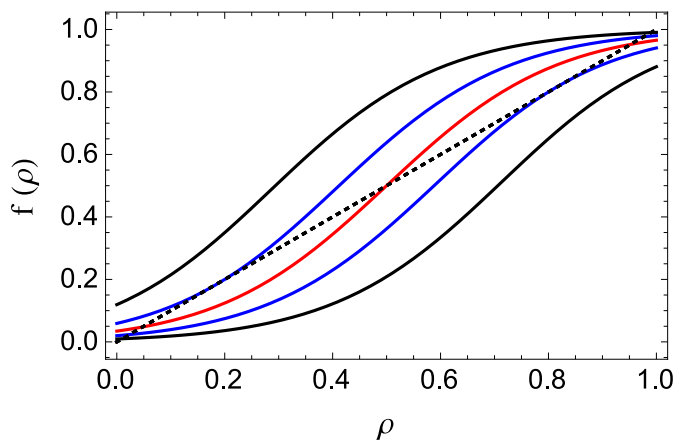}%
\caption{Plots of $f\left(  \rho\right)  $ for $\Delta<0$. We assume
temperature $k_{B}T/\epsilon=0.3$, ramification $r=5$, and several values of
the chemical potential, $\mu/\epsilon=-0.4$; $-0.17$; $0$; $0.17$; and $0.4$
(with ordered curves from left to right, for increasing values of $\mu$). The
black dotted line corresponds to $\rho$.
\label{map_first-order}}
\end{figure}

We first consider a simple case, $\epsilon_{aa}=\epsilon_{bb}=\epsilon/r>0$,
$\epsilon_{ab}=\epsilon_{ba}=0$, which corresponds to $I_{0}$ in the cubic
representation of Fantoni and collaborators \cite{fantoni13}. The attractors
of the map can be visualized if we draw graphs of $f\left(  \rho\right)  $,
given by Eq. (\ref{map-eq2}) in terms of $\rho$, as shown in figure \ref{map_first-order}, for a
fixed temperature ($k_{B}T/\epsilon=0.3$), a given ramification of the tree
($r=5$), and for several values of the chemical potential ($\mu/\epsilon
=-0.4$, $-0.17$, $0$, $0.17$ and $0.4$). Fixed points are solutions of the
equation $f(\rho)=\rho$, so that we plot $\rho$ as the black dotted line in
this figure. These plots cover all the three qualitatively distinct behaviors
of the map for $\Delta<0$. There can be a single stable fixed point (black
curves), a stable and a marginally stable fixed point (blue curves), and two
stable and one unstable fixed points (red curve). The plots suggest a
first-order transition from a high to a low density phase of $a$-particles.
The blue curves indicate the emergence (or vanishing) of two fixed points, as
well as their stability threshold. In a phase diagram in terms of temperature
and chemical potential, the blue curves describe the behavior of the map along
the spinodal lines. The red curve shows the behavior of the map at the
transition (in the next paragraph, we show that, below a certain critical
temperature, there is a first-order phase transition at $\mu=|\Delta
|/2-\delta$, with $\mu=0$ in the $I_{0}$ case of the cubic diagram). A
numerical inspection of Eqs. (\ref{map-eq}) and (\ref{map-eq2}) leads to no
additional characteristic structures of the map.%

\begin{figure}[!ht]
\centering
\includegraphics[width=0.6\linewidth]{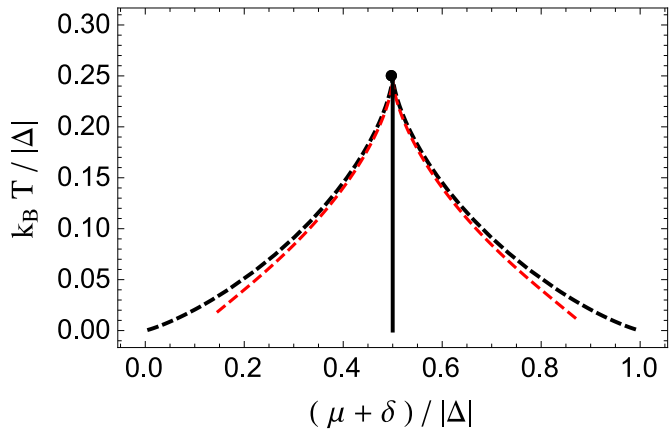}%
\caption{Phase diagram (temperature versus chemical potential) for $\Delta<0$.
The solid black line of first-order transitions ends at a critical point. We
also show spinodal lines (dashed) for a tree of finite coordination (red lines) and in
the mean-field limit (black lines). \label{first-diagram}}
\end{figure}

In figure \ref{first-diagram}, we draw some phase diagrams in terms of $(\mu+\delta)/|\Delta|$
and $k_{B}T/|\Delta|$ for a tree of finite coordination ($r=5$) and at the
infinite coordination limit. There is a critical point at $(\mu_{c}%
+\delta)/|\Delta|=1/2$, and $k_{B}T_{c}/|\Delta|=1/4$. The black solid line,
for $(\mu+\delta)/|\Delta|=1/2$, and $k_{B}T/|\Delta|<1/4$, is a first-order
boundary. The spinodal limits, which are represented by the dashed lines, are
given by the equations%
\begin{equation}
x_{0}=f(x_{0}),\quad|f(x_{0})|=1.
\label{spinodal-eq}
\end{equation}

\begin{figure}[!ht]
\centering
\includegraphics[width=0.6\linewidth]{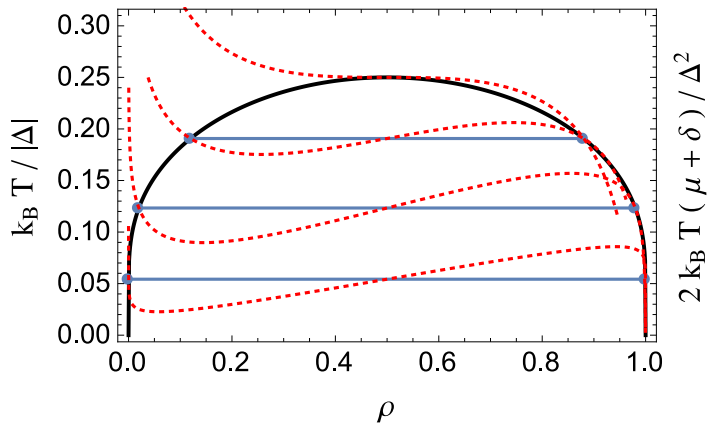}%
\caption{The black solid line is a coexistence curve (temperature versus
concentration of particles of type $a$). We also show tie lines (blue), and
isothems (red dotted lines) for the chemical potential in the mean-field
limit. \label{coexistence_plot}}
\end{figure}

In the infinite-coordination limit, we have the set of parametric equations%
\begin{equation}
\frac{k_{B}T}{\left\vert \Delta\right\vert }=\rho(1-\rho),
\end{equation}
and%
\begin{equation}
\frac{\left(  \mu+\delta\right)  }{\left\vert \Delta\right\vert }=\rho
(1-\rho)\ln\frac{1-\rho}{\rho}+\rho,
\end{equation}
with $0\leq\rho\leq1$. In order to describe the first-order boundary, we note
that, at a fixed point of $f_{\infty}$, we have%
\begin{equation}
(\mu+\delta)/|\Delta|=\rho+\frac{k_{B}T}{|\Delta|}\ln\frac{1-\rho}{\rho}.
\label{mu_T-rho}
\end{equation}
Therefore, $(\mu+\delta)/|\Delta|-1/2$ is an odd function of $\rho-1/2$, so
that
\begin{eqnarray}
\left\vert \int_{\rho_{1}}^{1/2}\left[  (\mu+\delta)/|\Delta|-1/2\right]
d\rho\right\vert 
%\nonumber \\ && \quad
=\left\vert \int_{1/2}^{1-\rho_{1}}\left[  (\mu
+\delta)/|\Delta|-1/2\right]  d\rho\right\vert ,
\end{eqnarray}
where $\rho_{1}$ is the smallest solution of $(\mu+\delta)/|\Delta|=1/2$. If
we resort to a Maxwell construction \cite{baxter82}, this leads to a
first-order boundary, given by $(\mu+\delta)/|\Delta|=1/2$. Along this border,
using Eq. (\ref{mu_T-rho}), we have%
\begin{equation}
\frac{k_{B}T}{|\Delta|}=\frac{1}{2}\frac{1-2\rho}{\ln\displaystyle\frac{1-\rho}{\rho}},
\label{coex_eq}
\end{equation}
so that the two coexistent densities are the solutions $(\rho,1-\rho)$ of Eq.
(\ref{coex_eq}), with $0\leq\rho\leq1$, and converge to $\rho\rightarrow1/2$ at
$T_{c}=|\Delta|/4k_{B}$. In figure \ref{coexistence_plot}, we show the coexistence curve (black
solid curve) and a few tie lines (in blue) in the mean-field limit. We also
show isotherms of the chemical potential (red dotted curves), which are
adequately scaled with temperature so that the corrected solutions, according
to Maxwell's construction, correspond to the coexistence tie lines. Similar
graphs and results can be numerically obtained for trees of finite
coordination (with $q\geq 2$). The cases $I_{0}$ and $B_{0}$ in the cubic
diagram of parameters are fully understood according to the general behavior
displayed in figures \ref{first-diagram} and \ref{coexistence_plot}.%

\section{Continuous transition ($\Delta > 0$)}

\begin{figure*}[th]
\begin{minipage}[b]{0.48\linewidth}
\centering
(a) \par\smallskip
\includegraphics[width=\linewidth]{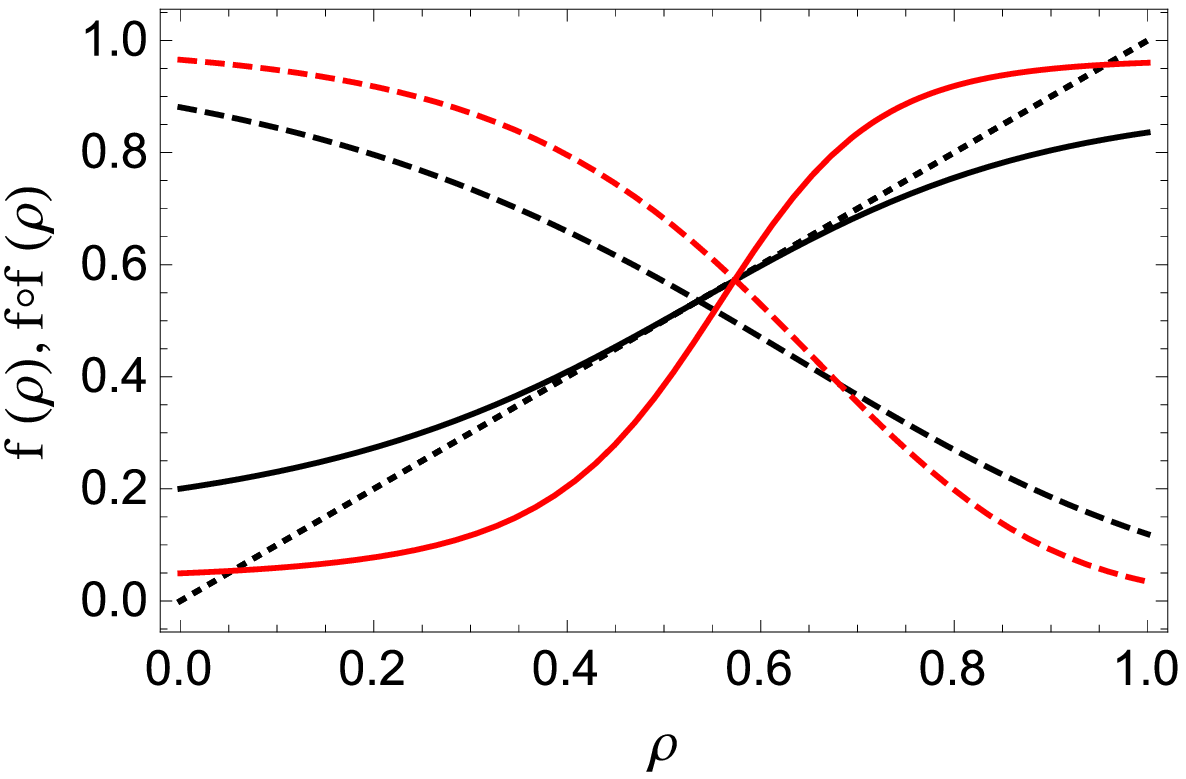}
\end{minipage}
\hspace{0.5cm}
\begin{minipage}[b]{0.48\linewidth}
\centering
(b) \par\smallskip
\includegraphics[width=\linewidth]{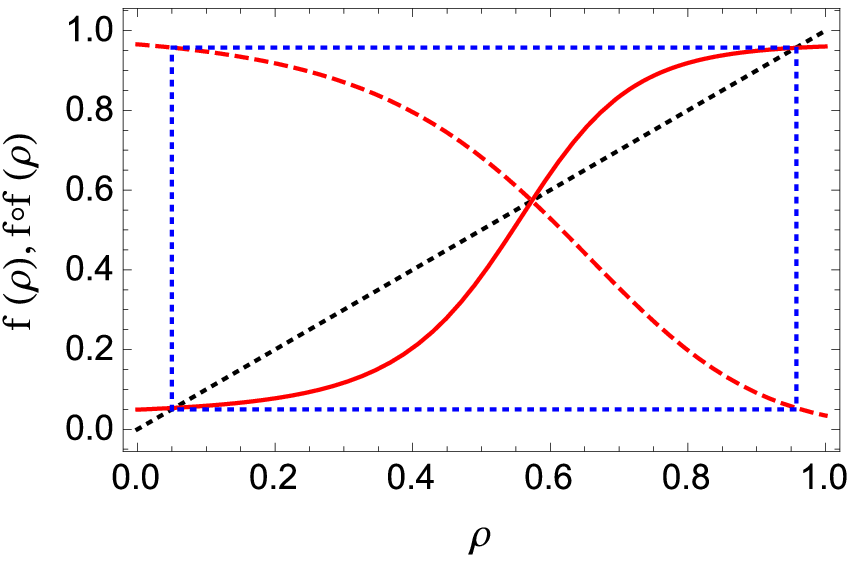}
\end{minipage}
\caption{ (a) Plots of $\rho$ (dotted line), $f(\rho)$ (dashed lines), and $f\circ
f(\rho)$ (solid lines), for fixed chemical potential, $\mu/\epsilon=0.5$,
ramification $r=7$, and two values of temperature, $k_{B}T/\epsilon=0.25$
(black) and $0.15$ (red). A cycle-$2$ orbit can be graphically found as the
solution of $\rho=f\circ f(\rho)$ with $\rho\neq f(\rho)$. (b) Cobweb plot to illustrate the period-$2$ cycle. \label{cycle2_maps}}%
\end{figure*}

For $\Delta>0$, at low temperatures, instead of a first-order boundary, there
is a critical line from a disordered to a cycle-$2$ periodic phase. We first
sketch the properties of the map in the simple $A_{0}$ case of the cubic
diagram of interactions, with $\epsilon_{ab}=\epsilon/r>0$, and $\epsilon
_{aa}=\epsilon_{bb}=\epsilon_{ba}=0$. Now there are only two scenarios. The
map has either a single stable fixed point or one unstable fixed point and a
stable cycle of period $2$. In figure \ref{cycle2_maps}a, we draw $\rho$ (dotted line),
$f(\rho)$ (dashed lines), and $f\circ f(\rho)$ (solid lines), for fixed
chemical potential, $\mu/\epsilon=0.5$, ramification $r=7 $, and two values of
temperature, $k_{B}T/\epsilon=0.25$ (black) and $0.15$ (red). A cycle-$2$
orbit can be graphically found as the solution of $\rho=f\circ f(\rho)$ with
$\rho\neq f(\rho)$. This orbit represents modulated phases, with density
oscillations along the generations of the tree (as illustrated in figure \ref{cayley}b).
The orbit of period $2$ can also be visualized by means of a cob-web plot,
which we show as the blue dotted line in figure \ref{cycle2_maps}b. Note that the instability
threshold occurs at the same point as the emergence of the stable cycle-$2$
phase, and that any fixed point located between the solutions of a stable
cycle-$2$ orbit will be unstable. A numerical inspection of $f$ and
$f_{\infty}$ leads to no additional structures of the map beyond these two scenarios.%

\begin{figure*}[th]
\begin{minipage}[b]{0.48\linewidth}
\centering
(a) \par\smallskip
\includegraphics[width=\linewidth]{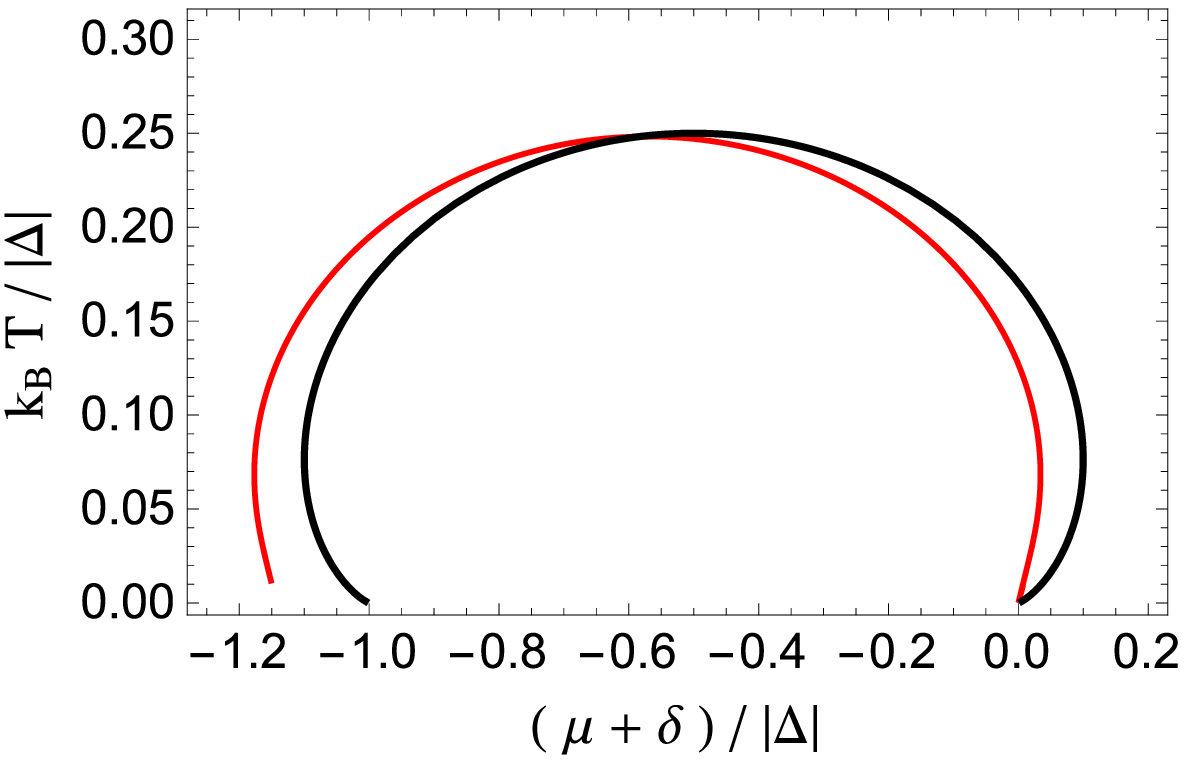}
\end{minipage}
\hspace{0.5cm}
\begin{minipage}[b]{0.48\linewidth}
\centering
(b) \par\smallskip
\includegraphics[width=\linewidth]{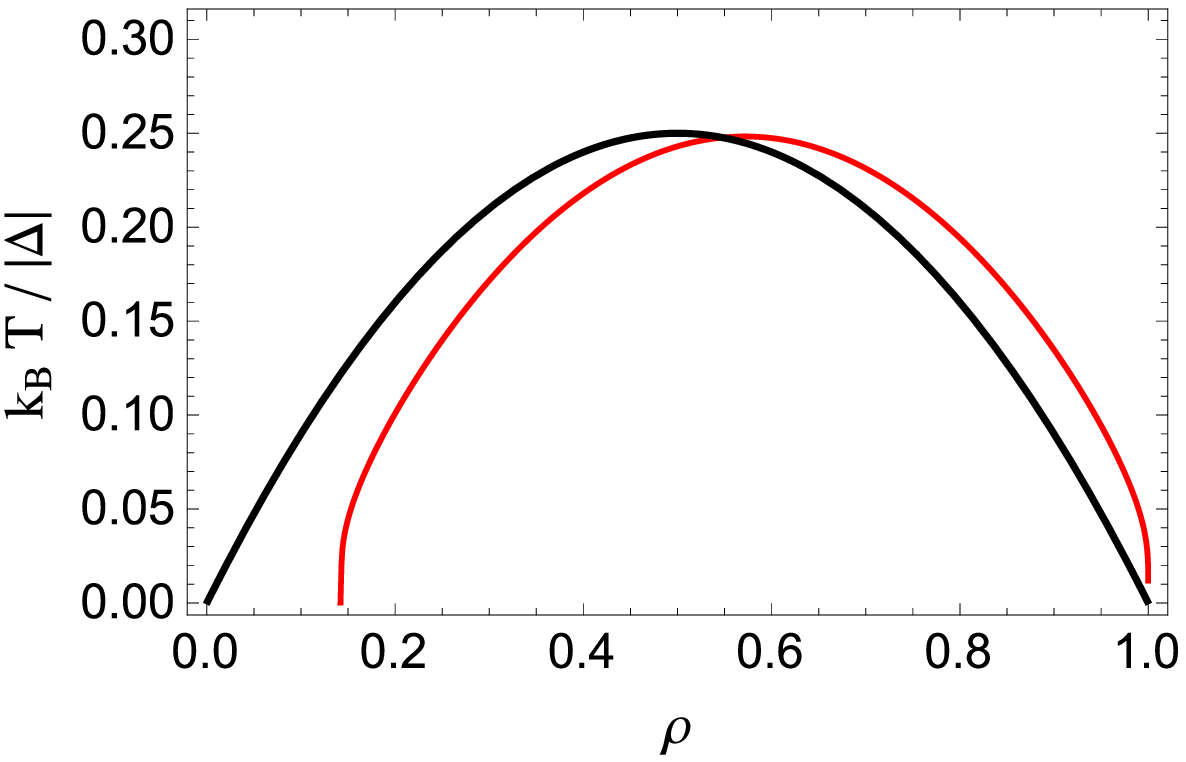}
\end{minipage}
\caption{ (a) Critical lines in the phase diagram in terms of temperature and
chemical potential (with $\Delta>0$). Red lines are for a tree of finite
coordination. Black lines are obtained in the infinite-coordination limit. There
is a modulated phase, of period $2$, inside the region enclosed by the
critical line. (b) Corresponding phase diagram in the $\rho \times T$ plane. \label{cycle2_diagram}}%
\end{figure*}

In figure \ref{cycle2_diagram}a we plot phase diagrams in terms of $(\mu+\delta)/|\Delta|$ and
$k_{B}T/|\Delta|$, with $\Delta>0$, in the mean-field limit (black line) and
for a tree of finite ramification (red lines), using the interactions
parameters of $A_{0}$ in the cubic diagram of Fantoni and collaborators
\cite{fantoni13}. Since there is a coincidence between stability and
transition thresholds, we can use Eq. (\ref{spinodal-eq}) to derive parametric
equations for the critical line of the mean-field map,%
\begin{equation}
\frac{k_{B}T}{|\Delta|}=\rho(1-\rho),
\end{equation}
and%
\begin{equation}
\frac{\mu+\delta}{|\Delta|}=\rho(1-\rho)\ln\frac{1-\rho}{\rho}-\rho.
\end{equation}
Note that the minus sign in the second term on the r.h.s. of this last equation is
responsible for both the \textquotedblleft soft\textquotedblright\ behavior
near $k_{B}T/|\Delta|=0.25$, and the reentrant behavior at low temperatures.
For trees of finite coordination, we resort to a simple numerical calculation
to find the analogous critical line. In figure \ref{cycle2_diagram}b we plot the corresponding phase diagram in terms of $\rho$ and $k_{B}T / |\Delta|$. As we mentioned above, there is no
indication of an alternative critical behavior in the region enclosed by the
critical line.%

\section{Duality and low coordination limit}

We have found an interesting (dual) relation between the solutions for a
cycle-$2$ orbit along a symmetry line and the phase-separated densities along
the coexistence curve. Consider the mean-field map with $\Delta>0$, so that at
$(\mu+\delta)/|\Delta|=-1/2$, and $k_{B}T/|\Delta|<1/4$, there is a periodic
solution of the form $(\rho,1-\rho)$. Thus, we can write%
\begin{equation}
1-\rho=f_{\infty}(\rho)=\left[  1+e^{-(1-\rho)/2\tilde{T}}\right]  ^{-1},
\end{equation}
from which we have%
\begin{equation}
\rho=\left[  1+e^{(1-\rho)/2\tilde{T}}\right]  ,\label{solution2}%
\end{equation}
where $\tilde{T}$ is a short-hand notation for $k_{B}T/|\Delta|$. Hence, the
solutions for the density along the coexistence curve for $\Delta<0$ are
identical to the periodic orbit solutions along the $(\mu+\delta
)/|\Delta|=-1/2$ symmetry line for $\Delta>0$.

We now consider the effects of the finite coordination of the tree on the
critical properties of the two-state Janus gas. Let us restrict the attention
to two representative cases of the cubic diagram of interactions, $I_{0}$ and
$A_{0}$, for $\Delta<0$ and $\Delta>0$, respectively. In the $I_{0}$ case, for
$(\mu+\delta)/|\Delta|=1/2$ and $\rho=1/2$, we have%
\begin{equation}
\left.  f^{\prime}(\rho)\right\vert _{\rho=1/2}=r\tanh\left(  \frac{|\Delta
|}{4\,r\,k_{B}\,T}\right)  ,
\end{equation}
so that, at the critical point,
\begin{equation}
\frac{1}{r}=\tanh\left(  \frac{|\Delta|}{4\,r\,k_{B}\,T_{c}}\right)  .
\label{finite-r_eq}
\end{equation}
As it should be anticipated, $T_{c}=|\Delta|/4k_{B}$ is a trivial solution for
the map in the mean-field limit ($r\rightarrow\infty$). Also, Eq.
(\ref{finite-r_eq}) has no solutions for $r<1$, so $T_{c}\rightarrow0$ as
$r\rightarrow1$. This result is consistent with the one-dimensional chain
structure of the Bethe lattice for $r=1$, since no transition is expected in
one-dimensional systems with short-range interactions. Similar results can be
obtained in the $A_{0}$ case (with $(\mu+\delta)/|\Delta|=-1/2$). We remark
that the symmetry line of the mean-field map is shifted to the left as $r$ is
decreased, and the solution $\rho=1/2$ at the critical point is no longer
valid for finite $r$ if we fix $(\mu+\delta)/|\Delta|=-1/2$. For finite
ramification, we have to numerically calculate the value of $\rho$ at the
critical temperature.

\section{Conclusions}

We have considered a lattice gas of two-state Janus particles on the sites of
a Cayley tree. Taking advantage of the geometrical structure of this graph, it
is particularly simple to introduce directional interactions between
first-neighbor sites along the branches of the tree. The problem is formulated
in terms of a set recursion relations, whose
attractors correspond to physical solutions on the Bethe lattice (the deep
interior of the Cayley tree). With relatively easy calculations, we can draw a
number of phase diagrams in terms of temperature and either chemical potential
or the concentration of a type of particles, and for a wide range of energy
parameters. We then make contact with recent simulations for the analogous
system of hard spheres with a short-range attractive potential well. The
results on the Bethe lattice provide a unified view of the systems represented
in a cubic diagram of interactions drawn by Fantoni and collaborators
\cite{fantoni13, maestre13}. In particular, depending on a
combination of energy parameters, which includes the analog of the directional
Kern-Frenkel potential, we show the existence of a critical line separating
disordered and cycle-$2$ modulated phases, which have been found in some
of the simulations for equal concentrations. The calculations on the Bethe
Lattice are simple enough to give qualitative results for all choices of
parameters of the model.

\section*{Acknowledgement}

We acknowledge the financial support provided by the Brazilian agencies CNPq and Fapesp.
% under Grants No. 05/04459-1 and
% 06/51286-8.

\appendix

\section{Derivation of recursion relations}

\begin{figure}[!ht]
\centering
\includegraphics[width=0.7\linewidth]{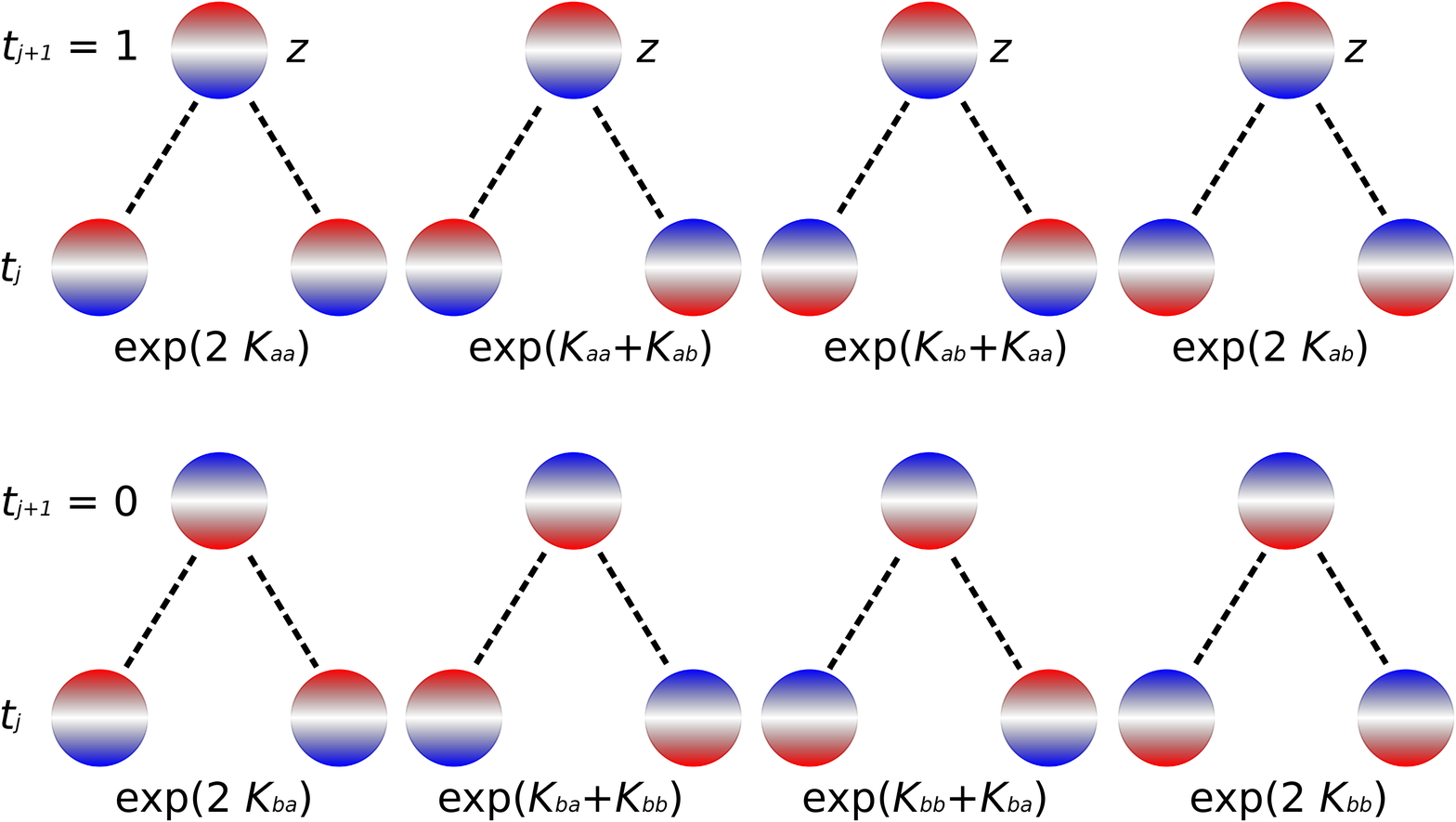}%
\caption{Illustrations of the possible configurations at the $j-th$ generation with $t_{j+1} =1$ (top) and $t_{j+1}=0$ (bottom). Note that we write the Boltzmann factor associated with each configuration. A fugacity term $z$ is present at the recursion relation if $t_{j+1}=1$. \label{cayley_recursion}}
\end{figure}

In figure \ref{cayley_recursion}, we draw the configurations that are associated with two generations of a Cayley tree of ramification $r=2$
(corresponding to coordination $q=r+1=3$). This tree is constructed along a
direction that leads to a consistent and unambiguous choice of the interaction
parameters. We assume that $a$ and $b$ particles interact with energy
$-\epsilon_{ab}$ if a particle of type $a$ is on a site belonging to a certain
generation and a particle of type $b$ is on a nearest-neighbor site belonging
to the next generation (and vice versa). There is a Boltzmann factor associated with each interaction between $j$ and $j+1$. There is also a fugacity term related to particles of type $a$ ($t_{j+1}=1$). For this $3$-coordinated tree, let
$\Xi_{j}^{a}$ ($\Xi_{j}^{b}$) be the partial grand partition function
associated with the sub-tree generated by a site at generation $j$ which is
occupied by a particle of type $a$ ($b$). Thus, the partial partition
functions at generations $j$ and $j+1$ obey the set of relations%
\begin{eqnarray}
\Xi_{j+1}^{a}=z\left[  e^{2K_{aa}}\left(  \Xi_{j}^{a}\right)  ^{2}%
+2e^{K_{aa}+K_{ab}}\left(  \Xi_{j}^{a}\Xi_{j}^{b}\right)  
%\right. \nonumber \\ && \quad \left.
+e^{2K_{ab}}\left(
\Xi_{j}^{b}\right)  ^{2}\right]
\end{eqnarray}
and%
\begin{eqnarray}
\Xi_{j+1}^{b}=e^{2K_{ba}}\left(  \Xi_{j}^{a}\right)  ^{2}+2e^{K_{ba}+K_{bb}%
}\left(  \Xi_{j}^{a}\Xi_{j}^{b}\right) 
%\nonumber \\ && \quad
+e^{2K_{bb}}\left(  \Xi_{j}%
^{b}\right)  ^{2},
\end{eqnarray}
where $K_{kl}=\beta\epsilon_{kl}$, for $k,l=a,b$, $\beta=1/k_{B}T$ is the
inverse temperature, $z=\exp(\beta\mu)$ is the fugacity, and $\mu$ is the
chemical potential (associated with particles of type $a$). For a tree with a
general ramification $r$, we have the more general equations%
\begin{equation}
\Xi_{j+1}^{a}=z\left[  e^{K_{aa}}\Xi_{j}^{a}+e^{K_{ab}}\Xi_{j}^{b}\right]
^{r}\label{xi-a_recur}%
\end{equation}
and%
\begin{equation}
\Xi_{j+1}^{b}=\left[  e^{K_{ba}}\Xi_{j}^{a}+e^{K_{bb}}\Xi_{j}^{b}\right]
^{r}.\label{xi-b_recur}%
\end{equation}

\section*{References}

\bibliographystyle{unsrt}

\end{document}